\documentclass[aps,prl,twocolumn,showpacs,showkeys,preprintnumbers,floats,amsmath,amssymb,floatfix]{revtex4-1}

\usepackage{graphicx}

\begin{document}

\preprint{arXiv}

\title{Overlapping-gate architecture for silicon Hall bar MOSFET devices\\ in the low electron density regime} 

\author{L.~H.~Willems~van~Beveren}
\email[Electronic mail: ]{l.h.willemsvanbeveren@unsw.edu.au}
\affiliation{Australian Research Council Centre of Excellence for Quantum Computer Technology,\\
School of Physics, The University of New South Wales, Sydney 2052, Australia}
\author{K.~Y.~Tan}
\affiliation{Australian Research Council Centre of Excellence for Quantum Computer Technology,\\
School of Electrical Engineering and Telecommunications, The University of New South Wales, Sydney 2052, Australia}
\author{N.~S.~Lai}
\affiliation{Australian Research Council Centre of Excellence for Quantum Computer Technology,\\
School of Electrical Engineering and Telecommunications, The University of New South Wales, Sydney 2052, Australia}
\author{A.~S.~Dzurak}
\affiliation{Australian Research Council Centre of Excellence for Quantum Computer Technology,\\
School of Electrical Engineering and Telecommunications, The University of New South Wales, Sydney 2052, Australia}
\author{A.~R.~Hamilton}
\affiliation{School of Physics, The University of New South Wales, Sydney 2052, Australia}

\date{\today}

\begin{abstract}
We report the fabrication and study of Hall bar MOSFET devices in which an overlapping-gate architecture allows four-terminal measurements of low-density 2D electron systems, while maintaining a high density at the ohmic contacts. Comparison with devices made using a standard single gate show that measurements can be performed at much lower densities and higher channel resistances, despite a reduced peak mobility. We also observe a voltage threshold shift which we attribute to negative oxide charge, injected during electron-beam lithography processing.
\end{abstract}



\maketitle


A common issue in low temperature measurements of enhancement-mode metal-oxide-semiconductor (MOS) field-effect transistors (FETs) in the low electron density regime is the high contact resistance dominating the device impedance. In that case a voltage bias applied across the source and drain contact of a Hall bar MOSFET will mostly fall across the contacts (and not across the channel) and therefore magnetotransport measurements become challenging. However, from a physical point of view, the study of MOSFET nanostructures in the low-electron density regime involves a number of interesting phenomena (impurity limited mobility~\cite{gold1988}, carrier interactions~\cite{dassarma1999,spivak2010} and spin-dependent transport~\cite{laurens2008}) and it is therefore important to come up with solutions that work around the problem of a high contact resistance.

Previously, a split-gate MOSFET technique~\cite{kopley1988,heemskerk1998} was developed with submicron gaps (50-70~nm) in the gate electrode which allowed one to maintain a high electron density in the vicinity of the contacts regardless of its value in the main part of the sample. This technique has permitted reliable measurements of two-dimensional transport at low densities in the quantum Hall regime~\cite{wang1991}. However, a prerequisite for this technique is that the gate oxide thickness must be larger than the gap size to ensure that the channel is continuous under the gap. Since it is challenging to fabricate in a reproducible manner narrow gaps on the nanometer scale over the full width of a MOSFET, this technique is not suitable for the study of MOSFET structures with very thin ($\sim$5~nm) gate dielectric. Moreover, the reactive ion etching process used to create the submicron gaps in the gate metallization could in principle reduce the device mobility.

In this Letter we present a simple device architecture that allows measurement of a thin-oxide Hall-bar MOSFET for very low electron densities in the channel, where the resistance of the contacts can be controlled electrically by separate electrodes (referred to as lead gates). The fabrication process involved, based on overlapping-gates, has been successfully demonstrated for the fabrication of tunable few-electron silicon quantum dots~\cite{angus2007,lim2009b} and does not require additional processing steps like atomic layer deposition (ALD) or polysilicon etch steps that are known to reduce the device mobility~\cite{nordberg2009}.
\begin{figure}[t!]
\includegraphics[width=7.0cm]{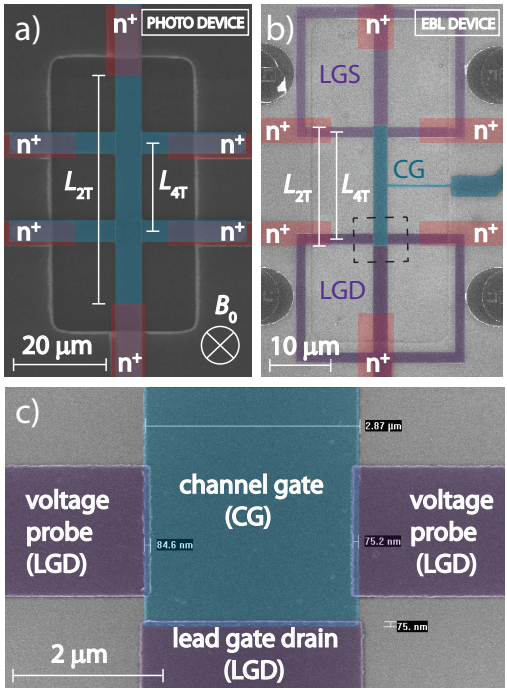}
\caption{(Color online) Scanning electron micrographs of the MOSFET device architectures studied. (a) Photolithography defined Hall bar with a single gate. (b) Electron-beam lithography defined Hall bar with overlapping-gate architecture. (c) Zoom-in of panel (b) showing the area where the channel gate (green) is electrically insulated from the lead gates (magenta) by a thin layer of aluminum oxide Al$_{x}$O$_{y}$.}
\label{fig:figure1}
\end{figure}

Figure~\ref{fig:figure1} shows the scanning-electron microscope (SEM) images of the two enhancement-mode Hall bar MOSFET device architectures studied in this work. The first device in panel (a) is fabricated by optical lithography (PHOTO) and has a channel dimension of 19.9 by 4.9~$\mu$m with $L/W$=4.06. Here a single aluminum gate (100~nm thickness) is patterned on top of a 5~nm SiO$_{2}$ gate dielectric which was grown by ultra-dry oxidation (UDOX) using dichloroethylene (DCE). The second device has a channel dimension of 19.0 by 2.87~$\mu$m, with $L/W$=6.62, and is fabricated by a two-step electron-beam lithography (EBL) process. The electron energy in the EBL process was 30~keV and a typical dose of 500-600~$\mu$C/cm$^{2}$ was used to expose the PMMA resist. The "channel gate" is defined by thermal aluminum evaporation (50~nm) and lift-off, followed by oxidation on a hotplate 150$^{\circ}$C to form a layer of aluminum oxide, with a thickness of a few nanometers~\cite{heij2001}. This dielectric film is used to electrically insulate the channel gate from a second layer of aluminum (100~nm), which defines "lead gates" to independently induce high-density electron layers connecting to the ohmic contacts. To avoid leakage (pinholes) between the two layers of aluminum the overlap at the contacts is kept to a minimum (about 80~nm by 2~$\mu$m). Both devices were subject to a final forming gas anneal (FGA) to reduce the interface trap density. The EBL device was also subject to a rapid thermal anneal (RTA) at 1000$^{\circ}$C for 5~seconds directly after the UDOX process. The oxide thickness $t_{ox}$ and interface trap density $D_{it}$ for both devices were independently measured on in-situ grown MOS capacitors by ellipsometry and CV-DLTS analysis~\cite{mccallum2008,johnson2010} to be $t_{ox}$=5.4$\pm$0.2~nm and $D_{it}$$\leq$1$\times$10$^{11}$~/eV/cm$^{2}$ (near the conduction band edge), respectively.

\begin{figure}[t!]
\includegraphics[width=7.0cm]{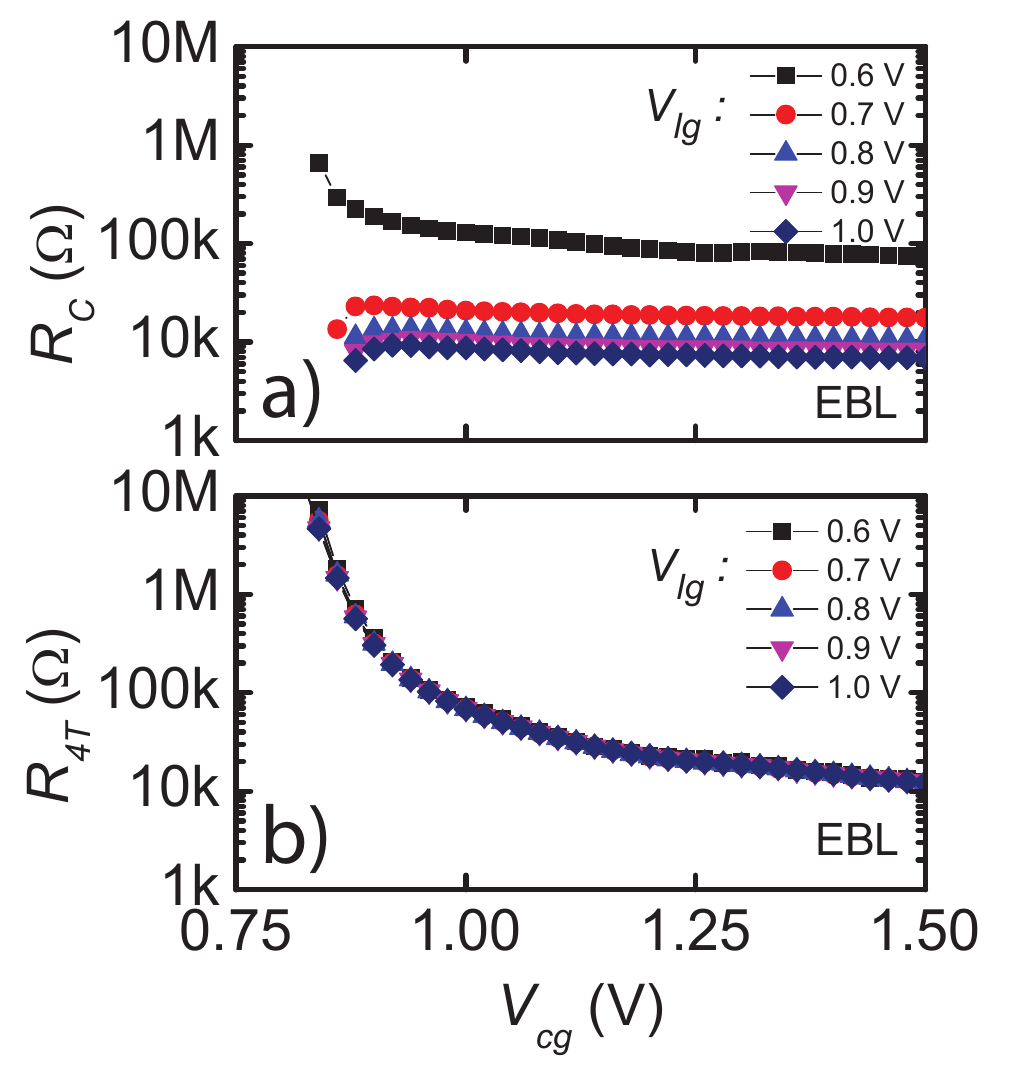}
\caption{(Color online) (a) Control of the contact resistance $R_{c}$ in the overlapping-gate Hall bar MOSFET (EBL fabricated device). Here only the channel gate voltage is swept as the voltage of the lead gates is increased stepwise from $V_{lg}$=0.6 to 1.0~V. (b) Four-terminal device resistance $R_{4T}$ as a function of the lead gate voltage.}
\label{fig:figure2}
\end{figure}

We now discuss the electrical transport characteristics of the two device architectures in detail. Magnetotransport measurements up to 8~T were performed in a dilution refrigerator containing a superconducting magnet with a base temperature of 20~mK, using standard 4-terminal AC lock-in techniques with an excitation voltage of 100-200~$\mu$V at 87~Hz.
Figure~\ref{fig:figure2}~(a) shows that the contact resistance $R_{c}$ of the Hall bar MOSFET with overlapping-gate geometry (EBL device) can be controlled, by adjusting the voltage of the lead gates $V_{lg}$, and is approximately independent from the channel gate voltage $V_{cg}$. The contact resistance was calculated by $R_{c}$=1/2($R_{2T}$-$\frac{L_{2T}}{L_{4T}}$$R_{4T}$), where $R_{2T}$ and $R_{4T}$ are the 2 and 4-terminal device resistances and $L_{2T}$=19.9~$\mu$m ($L_{4T}$=19.0~$\mu$m) is the length of the current trajectory from source to drain (channel), respectively. In the EBL device the source and drain contacts are much closer to the voltage probes than for the PHOTO device, $\frac{L_{2T}}{L_{4T}}$$\sim$~1. Figure~\ref{fig:figure2}~(b) shows the 4-terminal device resistance corresponding to each trace in Fig.~\ref{fig:figure2}~(a), demonstrating that $R_{4T}$ is independent of lead gate voltage. Even though the channel resistance $R_{4T}$ varies by three orders of magnitude, the contact resistance is essentially independent of the channel gate bias. For $V_{lg}$=1~V the contact resistance is always much less than the channel resistance. This is especially important for measurements at low carrier densities, where interaction effects are significant~\cite{kravchenko2000} but large $R_{c}$ makes it hard to cool the electrons~\cite{prus2001}.

\begin{figure}[t!]
\includegraphics[width=7.0cm]{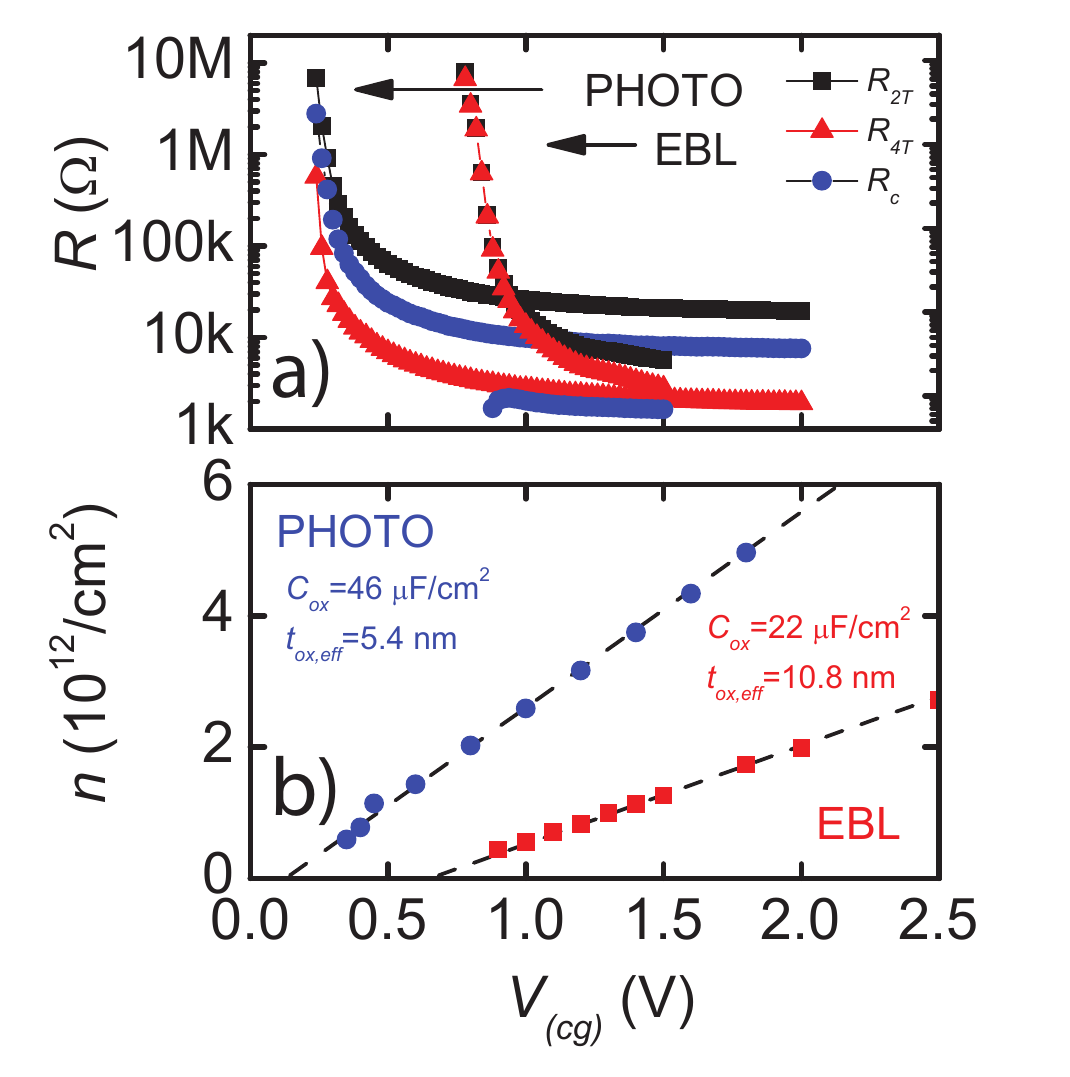}
\caption{(Color online) Device resistance (a) and electron density (b) as function of (channel) gate voltage for the two device architectures studied.}
\label{fig:figure3}
\end{figure}

The device resistance of the two device architectures as a function of applied channel gate voltage is compared in Fig.~\ref{fig:figure3}~(a). For the PHOTO device the contact resistance is always dominating the channel resistance ($R_{c}\gg R_{4T}$). This is particularly evident close to threshold, where $R_{2T}$ is starting to get very large ($>$10~M$\Omega$), even though $R_{4T}$ is only 1~M$\Omega$. In contrast, in the EBL device, with lead-gates set to $V_{lg}$=1~V, we are able to keep the carrier density near the ohmic contacts high, so that $R_{c}$ is always less than $R_{2T}$. This enables us to measure reliably to much larger channel resistances $R_{4T}$$>$100~M$\Omega$, limited only by the input impedance of the voltage pre-amplifier (lock-in) used. Measurements with high impedance voltage preamplifiers will be carried out in the future to probe this regime in more detail. For the calculation of the contact resistance of the PHOTO device we used $L_{2T}$=45.3~$\mu$m and $L_{4T}$=19.9~$\mu$m, so that $\frac{L_{2T}}{L_{4T}}$=2.27.

Additionally, from the Hall effect measurements it is possible to obtain the two-dimensional electron gas (2DEG) density $n$ as a function of (channel) gate voltage $V_{(cg)}$. From Fig.~\ref{fig:figure3}~(b) we can extract the gate capacitance per area of the two Hall bar devices $\frac{C_{ox}}{A}$=$\frac{\epsilon_{0}\epsilon_{r,\textrm{eff}}}{t_{ox,\textrm{eff}}}$=$\frac{Q}{V}$=e$\frac{dn}{dV}$, where $A$ is the channel area, $t_{ox,\textrm{eff}}$ is the effective SiO$_{2}$ thickness and $\epsilon_{0}$ ($\epsilon_{r,\textrm{eff}}$) is the (effective relative) dielectric constant, respectively. Since the 2DEG is formed within 10~nm of the silicon crystal and the oxide film is only 5.4~nm thick, we use an effective dielectric constant defined by 1/$\epsilon_{r,\textrm{eff}}$=1/$\epsilon_{r}$(Si)+1/$\epsilon_{r}$(SiO$_{2}$) resulting in $\epsilon_{r,\textrm{eff}}$=2.82 using $\epsilon_{r}$(Si)=11.9 and $\epsilon_{r}$(SiO$_{2}$)=3.7. For the PHOTO device, using $C_{ox}/A$=46~$\mu$F~/cm$^{2}$ as obtained from Fig.~\ref{fig:figure3}~(b), we extract $t_{ox,\textrm{eff}}$=5.4~nm, in excellent agreement with CV-DLTS measurements~\cite{mccallum2008,johnson2010}. However, the slope of $n(V)$ is noticeably different for the EBL device, despite both devices having identical SiO$_{2}$ thicknesses. The difference in slopes (gate capacitance) indicates a difference in gate dielectric between the devices. Cross-sectional transmission electron microscopy (X-TEM) studies on similar devices has shown that the Al oxidation process, used to form the overlapping gates, leads to an extra insulating layer of Al$_{x}$O$_{y}$ at the SiO$_{2}$/Al interface~\cite{lim2009a}. This film has a dielectric constant $\epsilon_{r}$(Al$_{2}$O$_{3}$)=11.5 (sapphire) that is more than twice the value for SiO$_{2}$. This results in a lower $C_{ox}$ and an apparent thicker $t_{ox}$ if we assume only SiO$_{2}$ is present between the gate and the channel. From this data we extract a sapphire thickness of $t_{ox}$(Al$_{x}$O$_{y}$) $\sim$2~nm, in reasonable agreement with the previous X-TEM study. The second key difference between the two device architectures is the large shift in threshold voltage $\Delta V_{T}$, as seen in Fig.~\ref{fig:figure3}~(a), which we will return to shortly. By plotting $n$ versus $V/t_{ox}$ we estimate that a total negative charge of $\sim$10$^{12}$~cm$^{-2}$ is responsible for the $\sim$400~mV shift in threshold voltage.

\begin{figure}[t!]
\includegraphics[width=7.0cm]{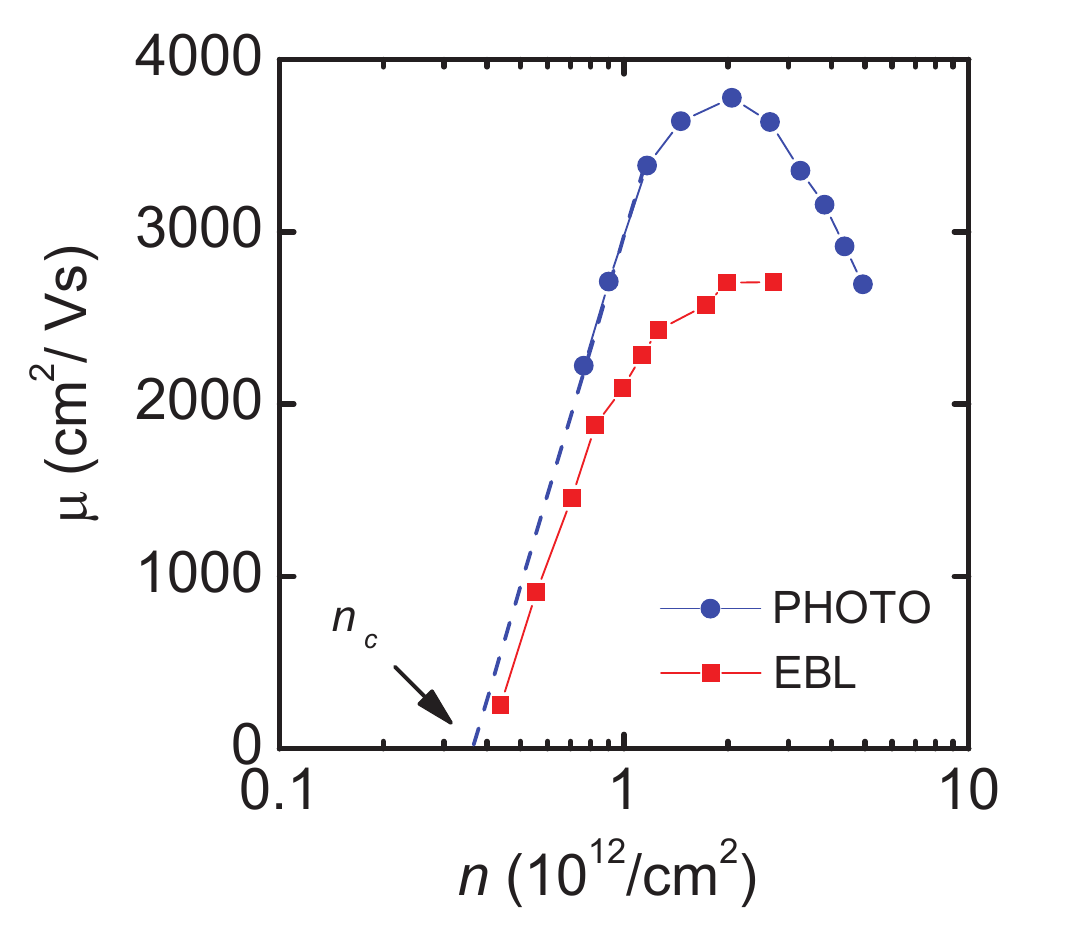}
\caption{(Color online) Mobility versus electron density. Note that with the EBL device one can measure at much lower densities as compared to the PHOTO device.}
\label{fig:figure4}
\end{figure}

The channel mobility $\mu$ was measured as function of electron density $n$  and is compared between the two types of devices in Fig.~\ref{fig:figure4}. The data demonstrate that the EBL device allows precise mobility measurements at lower 2DEG densities than possible with the PHOTO device.
In the low electron density regime the critical density $n_{c}$ (where $\mu$=0) measured on the PHOTO device extrapolates (blue dashed line) to the same value as measured with the EBL device, $n_{c}$$\sim$4$\times$10$^{11}$/cm$^{2}$. Equating the critical density with the interface trap density~\cite{dassarma1999}, this suggests that the interface trap density $D_{it}$ is the same for both device architectures, corresponding to an average area of $\sim$250~nm$^{2}$ per trap. In the high electron density regime, interface roughness limits the mobility. The larger gate voltages applied in this regime pull the electron wavefunction closer to the Si/SiO$_{2}$ interface. In the intermediate electron density regime the peak mobility is determined by the interplay of impurity scattering and interface roughness. In this regime we observe a peak mobility for the PHOTO device of $\mu$$\sim$3800~cm$^{2}$/Vs at a density of $n$=2$\times$10$^{12}$/cm$^{2}$ which is consistent with previous calculations~\cite{gold1985} and experiments~\cite{mccamey2005}. In comparison, for the EBL device the peak mobility is substantially reduced to $\mu$$\sim$2700~cm$^{2}$/Vs. If we assume $D_{it}$ is the same for both device types, we can only conclude that the large shift in threshold voltage $V_{T}$ observed is related to negative fixed oxide charge, arising from the EBL device processing. This additional charge is a possible explanation for the reduction in peak mobility for the EBL device. Previous studies have showed that EBL processing (even after a post-processing FGA) can cause threshold shifts of up to $\sim$400~mV due to negative oxide charge, screening the gate electrode~\cite{ma1989}. We estimated from Fig.~\ref{fig:figure3}~(b) the induced charge to be $\sim$10$^{12}$/cm$^{2}$. The trapped charge density and threshold shift are consistent with results of Aitken~\cite{aitken1979}.

In summary, we have shown that the overlapping-gate device architecture allows accurate mobility measurements in the low electron density regime, not limited by contact resistance. The (extrapolated) critical density, or interface trap density, is the same for the two device architectures. We observe a large threshold voltage shift for the EBL device as compared to the PHOTO device. The EBL processing reduces the peak mobility in the intermediate electron density regime in comparison to the PHOTO device, possibly due to an additional scattering mechanism. The fact that the mobility for both device architectures in the low electron density regime is similar provides further evidence that the threshold shift is caused by fixed oxide charge and not by interface traps.

The authors wish to thank D. Barber for assistance and acknowledge support from the Australian Research Council (ARC), the Australian Government, the U.S. National Security Agency and the U.S. Army Research Office under Contract No. W911NF-08-1-0527. A.R.H. acknowledges support from an ARC APF grant No. DP0772946.


\end{document}